\documentclass[journal=jacsat,manuscript=article]{achemso}

\usepackage[version=3]{mhchem} 
\usepackage{xcolor}

\newcommand{\dd}{\text{d}}
\newcommand*{\kb}{k_{\rm{B}}}

\author{Serhii Volosheniuk}
\email{s.volosheniuk@tudelft.nl}
\affiliation{Kavli Institute of Nanoscience, Delft University of Technology, Lorentzweg 1, Delft, 2628 CJ, The Netherlands}
\author{Riccardo Conte}
\affiliation{Kavli Institute of Nanoscience, Delft University of Technology, Lorentzweg 1, Delft, 2628 CJ, The Netherlands}

\author{Eugenia Pyurbeeva}
\affiliation{The Fritz Haber Center for Theoretical Chemistry, Institute of Chemistry, The Hebrew University of Jerusalem, Jerusalem 9190401, Israel}

\author{Thomas Baum}
\affiliation{Kavli Institute of Nanoscience, Delft University of Technology, Lorentzweg 1, Delft, 2628 CJ, The Netherlands}

\author{Manuel Vilas-Varela}
\affiliation{Centro Singular de Investigaci\'{o}n en Qu\'{i}mica Biol\'{o}xica e Materiais Moleculares (CiQUS) and Departamento de Química Orgánica, Universidade de Santiago de Compostela, Santiago de Compostela, 15782, Spain}

\author{Saleta Fern\'{a}ndez}
\affiliation{Centro Singular de Investigaci\'{o}n en Qu\'{i}mica Biol\'{o}xica e Materiais Moleculares (CiQUS) and Departamento de Química Orgánica, Universidade de Santiago de Compostela, Santiago de Compostela, 15782, Spain}

\author{Diego Pe\~{n}a}
\affiliation{Centro Singular de Investigaci\'{o}n en Qu\'{i}mica Biol\'{o}xica e Materiais Moleculares (CiQUS) and Departamento de Química Orgánica, Universidade de Santiago de Compostela, Santiago de Compostela, 15782, Spain}
\alsoaffiliation{Oportunius, Galician Innovation Agency (GAIN), Santiago de Compostela 15702, Spain}

\author{Herre S.J. van der Zant}
\affiliation{Kavli Institute of Nanoscience, Delft University of Technology, Lorentzweg 1, Delft, 2628 CJ, The Netherlands}
\email{h.s.j.vanderzant@tudelft.nl}

\author{Pascal Gehring}
\email{pascal.gehring@uclouvain.be}
\affiliation{IMCN/NAPS, Universit\'{e} catholique de Louvain, Pl. de l'Universit\'{e} 1, Louvain-la-Neuve, 1348, Belgium}

\title[An \textsf{achemso} demo]
  {A Single-Molecule Quantum Heat Engine }

\abbreviations{EMBJ, SME-2OS, DCM, MCBJ, STM, CDP, DC, AC, FWHM,}

\begin{document}



\begin{abstract}
Particle-exchange heat engines operate without moving parts or time-dependent driving, relying solely on static energy-selective transport. Here, we realize a particle-exchange quantum heat engine based on a single diradical molecule, only a few nanometers in size. We experimentally investigate its operation at low temperatures and demonstrate that both the power output and efficiency are significantly enhanced by Kondo correlations, reaching up to 53\% of the Curzon–Ahlborn limit. These results establish molecular-scale particle-exchange engines as promising candidates for low-temperature applications where extreme miniaturization and energy efficiency are paramount.

\end{abstract}

\section{Introduction}
A heat engine converts part of the heat flowing from a hot to a cold reservoir into work (Fig. \ref{fig:Deviceandmeasurements}a). Classical heat engines exist across a broad range of scales, from macroscopic systems such as steam engines to microscopic devices, including micro-electromechanical~\cite{Wha03}, piezoresistive~\cite{Ste11}, and single-atom engines~\cite{Ros16}. Quantum heat engines exploit discrete energy levels to perform work while interacting with their environment via heat or particle exchange (see Fig. \ref{fig:Deviceandmeasurements}b). These systems must be open and their working medium -- analogous to the gas or liquid in classical engines -- can consist of quantum particles~\cite{Qua09}, spin systems~\cite{Fel03}, harmonic oscillators~\cite{Rez06}, or single electrons. Furthermore, particle exchange quantum heat engines are autonomous and operate continuous without time-dependent external driving. 

A prominent example for particle exchange quantum heat engines which relies on single-electron exchange (see Fig. \ref{fig:Deviceandmeasurements}c), is the quantum dot (QD) heat engine. Its working principle is based on energy filtering, where efficiency is maximized if particle exchange occurs through a narrow energy window -- smaller than the thermal energy, \( k_B T \) -- to suppress entropy generation during transport~\cite{Mahan96}. Experimentally, quantum dots formed in nanowires, tunnel-coupled to two electron reservoirs, have been demonstrated to be precise energy filters, achieving efficiencies close to the Curzon–Ahlborn limit~\cite{Josefsson2018}. Therein, key parameters influencing efficiency are the load resistance\cite{Josefsson2019} and tunnel coupling, which governs charge transport and energy selectivity\cite{Sachin}. Recent studies further emphasize the role of quantum dot degeneracies, which are directly linked to entropy generation\cite{Pyurbeevananolett.1c03591,Pyurbeeva22}, and strong electronic correlations~\cite{svilans18,Hsu22,Volosheniuk2025}, in optimizing heat-to-work conversion.

Extending this concept, single molecules have emerged as promising candidates for quantum heat engines. Their unique advantage lies in the ability to chemically tailor their functionality at the molecular level, allowing precise control over the position and width of the effective energy filter, which facilitates the energy conversion process. Unlike inorganic quantum dots, molecules offer additional tunability through their electronic degeneracies, spin ground states, coupling strengths, and internal interference effects which can be engineered to enhance thermoelectric efficiency~\cite{D5TA01503K,PhysRevB.101.045410}.  

In this work, we explore how strong electron correlations, particularly those giving rise to the Kondo effect, impact the efficiency of a single-molecule heat engine.
By performing single-molecule thermoelectric experiments under different load resistors and tuning the intra-molecular electron interactions with a magnetic field, we show that the maximum power output of the engine is significantly enhanced in the presence of Kondo correlations. We quantify this enhancement by introducing an asymmetry parameter, $\sigma$, which we extract from fits to the experimental data. We show that $\sigma$ is a reduced characteristic of the internal dynamics of the QD responsible for its thermoelectric performance. The understanding obtained in this study is crucial for optimizing heat-to-work conversion in molecular-scale thermoelectric devices, where many-body interactions can significantly alter transport properties. By leveraging molecular design strategies and correlation effects, it will be possible to refine energy filtering mechanisms and push the efficiency of nanoscale heat engines to the limits.


\section{Main}
\begin{figure*}[h!]
    \includegraphics[width=1.0\textwidth]{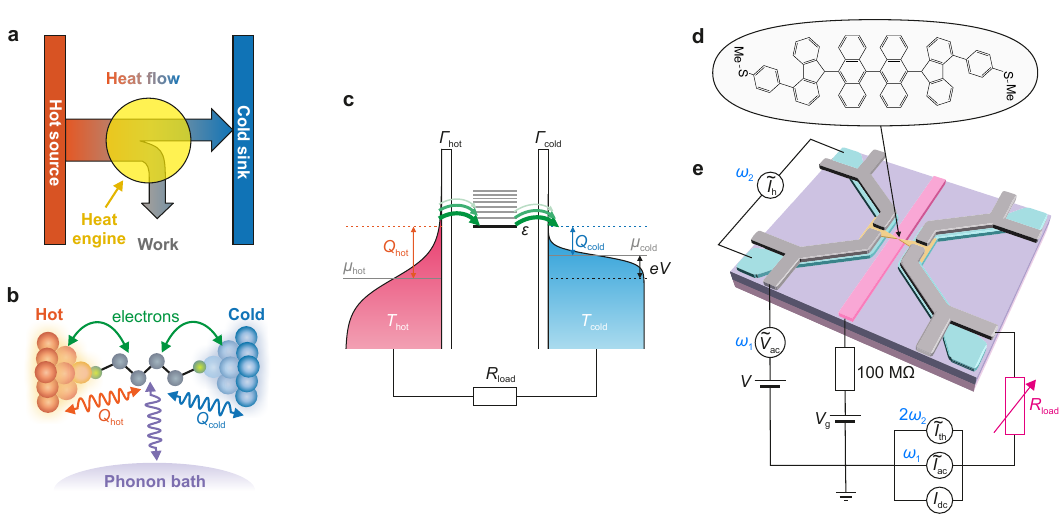}
    \caption{\label{fig:Deviceandmeasurements} (a) Working principle of a classical heat engine. (b) Principle of a molecular quantum heat engine. Electrons (green arrows) and heat (orange/blue/purple wiggly arrows) are exchanged between the molecule and a hot and cold reservoir. The molecule can furthermore thermalize with its phonon bath. (c) Molecular particle-exchange heat engine. A single molecule (depicted by a ground state in black and (vibrational) excited states in grey) is tunnel-coupled (green arrows) to a hot (left, red) and a cold (right, blue) particle/heat reservoir characterized by Fermi-Dirac distributions (with temperatures $T_\mathrm{hot}$, $T_\mathrm{cold}$ and electrochemical potentials $\mu_\mathrm{hot}$, $\mu_\mathrm{cold}$). An electron, driven by the thermal bias, tunnelling onto the molecular ground state removes a heat, $Q_\mathrm{hot}$, from the hot reservoir, converts part of it into useful work $eV$ -- which can be optimized by adjusting a load resistor, $R_\mathrm{load}$ --  and transfers the remaining heat, $Q_\mathrm{cold}$, to the cold reservoir. (d) Structure of the \textbf{SMe-2OS} molecule used in this study. (e) Schematic of the thermopower device showing the back-gate electrode (pink), heaters (light blue), gold bridge (yellow), and gold contacts (grey). The accompanying circuit diagram indicates the terminals used to apply the gate voltage, $V_\mathrm{g}$, dc bias voltage, $V$, ac bias voltage at frequency $\omega_1$, $V_\mathrm{ac}$~($\omega_1$), and ac heater current at frequency $\omega_2$, $I_\mathrm{h}$~($\omega_2$), as well as the terminals for measuring the dc current, $I_\mathrm{dc}$, ac current, $I_\mathrm{ac}$~($\omega_1$), and thermocurrent, $I_\mathrm{th}$~($\omega_2$). A load resistor, $R_\mathrm{load}$, is connected in series with the molecular heat engine.
 }
    
\end{figure*}

The working principle of the single-molecule particle-exchange heat engine studied in this work is illustrated in Fig.~\ref{fig:Deviceandmeasurements}b and c. A single molecule, characterized by its discrete energy levels --specifically, the highest occupied molecular orbital (HOMO) at energy $\varepsilon_{\text{HOMO}}$ and the lowest unoccupied molecular orbital (LUMO) at energy $\varepsilon_{\text{LUMO}}$ -- is tunnel-coupled to two electron reservoirs with electrochemical potentials $\mu_\mathrm{hot}$ and $\mu_\mathrm{cold}$ on the left (hot) and right (cold) sides, respectively. The tunnel coupling strengths to these reservoirs are denoted as $\Gamma_\mathrm{hot}$ and $\Gamma_\mathrm{cold}$.  

Typically, in molecular devices the energy gap $\varepsilon_{\text{HOMO}} - \varepsilon_{\text{LUMO}}$ is several eV, allowing transport to be effectively described as transitions through the single energy level closest to the Fermi energy (single-level model). The energy required to add an additional electron to this level is given by $\varepsilon=E_{\text{HOMO}}-E_{\text{LUMO}}-\mu$ (where $\mu$ is the chemical potential of the electrode), which accounts for both the charging energy and the quantum level spacings \cite{Gehring2019_theory, Pyurbeeva2020b}. When a temperature bias, $\Delta T = T_\mathrm{hot} - T_\mathrm{cold}$, is applied across the molecule, an electronic energy (heat) current $J_Q$ -- the phononic heat current can typically be neglected at cryogenic temperatures that are much lower than the vibrational energies of the single molecule -- flows from the hot to the cold reservoir, filtered by the molecular level at $\varepsilon$. Simultaneously, this thermal gradient drives a thermoelectric current, $I_{\text{th}}$, against the applied voltage, $V = (\mu_\mathrm{cold} - \mu_\mathrm{hot})/e$. Each transferred electron performs electrical work, $eV$, while transporting heat, extracting $Q_\mathrm{hot} = \varepsilon - \mu_\mathrm{hot}$ from the hot reservoir and depositing $Q_\mathrm{cold} = \varepsilon - \mu_\mathrm{cold}$ into the cold reservoir. If a load resistor $R_{\text{load}}$ is connected in series with the molecular junction, electrical power, $P = I_{\text{th}}^2 R_{\text{load}}$, is generated. The efficiency of the heat engine is then given by $\eta = \frac{P}{J_Q}.$  

In our experiment, we implement a single-molecule particle-exchange heat engine by incorporating an all-organic \textbf{SMe-2OS} molecule (Fig. \ref{fig:Deviceandmeasurements}d) -- a stable diradical which was previously synthesized and studied on surface by Scanning Tunneling Microscopy (STM)~\cite{STM_2OS} (see Methods) -- into a nanometer-sized junction between two gold electrodes. Here, the thioether SMe(-$\mathrm{SCH_3}$) end groups are included for anchoring to the Au electrodes. Notably, the molecule's spin ground state can be tuned to $S=1/2$ by applying a negative gate voltage, $V_\mathrm{g}$, as we will demonstrate later. This junction is formed through the electromigration~\cite{Oneil_migration} of a narrow gold constriction (see Methods). Our recently developed device architecture~\cite{Gehring2021,Volo23} (Fig.~\ref{fig:Deviceandmeasurements}e) allows for the application and precise measurement of a thermal bias across the junction, even at millikelvin sample temperatures. By simultaneously applying an AC bias voltage, \( V_{\mathrm{ac}} \), at frequency \( \omega_1 \), and an AC heater current, \( \tilde{I}_\mathrm{H} \), at \( \omega_2 \), we simultaneously measure both the electrical differential conductance, \( \mathrm{d}I/\mathrm{d}V = \tilde{I}_\mathrm{ac}/V_\mathrm{ac} \), at \( \omega_1 \), and the thermocurrent, \( I_\mathrm{th} \), at \( \omega_2 \) as a function of DC bias voltage, $V$, and $V_\mathrm{g}$ (see Methods for details).

\begin{figure*}[h!]
    \includegraphics[width=\textwidth]{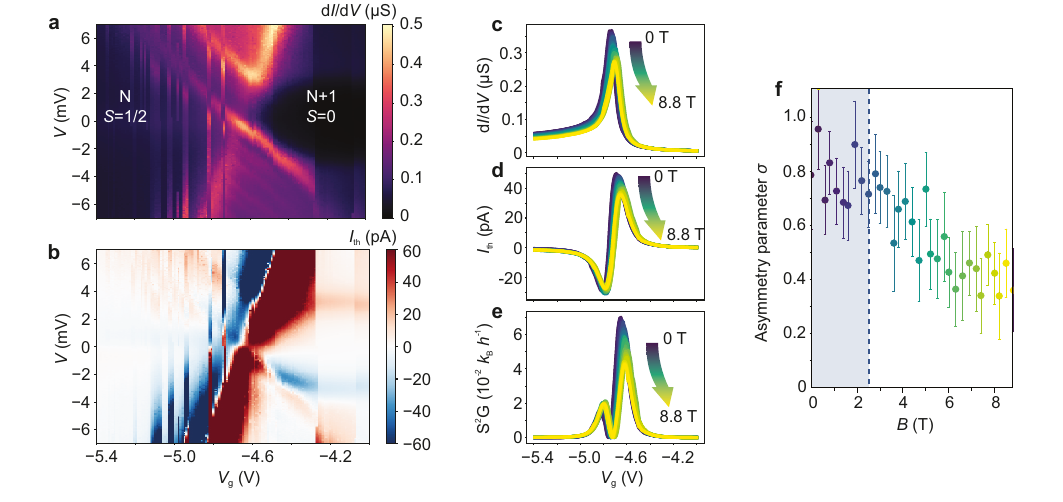}
    \caption{\label{fig2} 
    Simultaneously measured stability diagrams of differential conductance (a) and thermocurrent (b). Zero-bias gate traces of differential conductance (c) and thermocurrent (d) at various magnetic fields, showing a shift of the charge degeneracy point toward less negative gate voltages. (e) Evolution of the Power Factor calculated from (c) and (d) as a function of gate voltage for different magnetic fields. (f)  Asymmetry parameter extracted from thermopower measurements as a function of applied magnetic field.  Magnetic field strengths are represented by color: dark blue (0 T) transitioning to yellow (8.8 T). }
    
\end{figure*}

Figure \ref{fig2}a and b show maps of \( \mathrm{d}I/\mathrm{d}V \) and \( I_\mathrm{th} \) as a function of \( V \) and \( V_\mathrm{g} \). The characteristic hourglass-shaped sequential tunneling regime is observed, a hallmark of Coulomb blockade, with a charge degeneracy point (CDP) at \( V_\mathrm{g} = -4.7 \) V, marking the transition where one electron is removed from the molecule, shifting the system from an \( N+1 \) to an \( N \) charge ground state. Excited states appear within the sequential tunneling regime and extend into the Coulomb-blocked region of the \( N+1 \) state as horizontal lines, characteristic of second-order co-tunneling processes. The response of those features to an external magnetic field indicates that they originate from spin-excited states rather than vibrational excitations. Magnetic field dependence (see SI) reveals that the two spins in the \( N+1 \) charge state couple antiferromagnetically, with an exchange coupling strength of \( J \approx 3 \)~meV, confirming that the observed co-tunneling excitation corresponds to a singlet-triplet transition. Additionally, the CDP shifts to less negative energies under an applied magnetic field (see Fig. \ref{fig2}c, also SI), indicating that the ground state on the left side of the CDP has a higher spin than on the right. This implies a doublet-to-singlet transition upon electron addition with the neutral state of the molecule on the right hand side of the CDP.  

In the \( N \) charge ground state -- with doublet spin ground state -- a horizontal feature at zero bias is visible, which we attribute to the Kondo effect -- a many-body phenomenon where a localized spin on the molecule interacts with the conduction electrons of the reservoirs, forming an anti-ferromagnetically ordered bound state with a binding energy characterized by the Kondo temperature, $T_\mathrm{K}$. 
Magnetic field dependent measurements of $I_\mathrm{th}$ can be used to estimate \( T_\mathrm{K} \)~\cite{Hsu22}: A universal scaling relation exists between the magnetic field $B_\mathrm{c}$, where the Kondo resonance splits and $B_\mathrm{th}$ where the slope of $I_\mathrm{th}$ at zero bias changes sign, with $B_\mathrm{th} \approx B_\mathrm{c} = 0.75 k_\mathrm{B}T_\mathrm{K} / (g\mu_\mathrm{B})$, where $g$ is the g-factor and $k_\mathrm{B}$ is the Boltzmann constant.~\cite{Hsu22} We find (see SI)  $B_\mathrm{th} = 2.4$~T, yielding a Kondo temperature of \( T_\mathrm{K} = 4.3 \)~K. Furthermore, magnetic field sweeps at \( V_\mathrm{g} = -5.6 \)~V indicate that the system behaves as a spin-\( 1/2 \) with a \( g \)-factor of 2.

In Fig.~\ref{fig2}c-e, we present the gate-dependent \( \mathrm{d}I/\mathrm{d}V \), \( I_\mathrm{th} \) and the thermoelectric power factor \( PF = S^2G = I_\mathrm{th}^2 / \Delta T^2 / (\mathrm{d}I/\mathrm{d}V) \) at various magnetic field strengths, measured with a series resistor of \( R_\mathrm{load} = 100 \)~k\( \Omega \) at zero bias voltage. We observe that all three quantities are suppressed as the magnetic field increases. Additionally, the ratio between the maximum (\( I_{\mathrm{th}+} \)) and minimum (\( I_{\mathrm{th}-} \)) thermocurrent, decreases from 1.66 at zero field to 1.3 at 8.8~T. This reduction enhances the symmetry of the gate-dependent \( PF \) with respect to the charge degeneracy point (CDP).

To quantify the asymmetry of the thermoelectric response around the charge degeneracy point (CDP), we introduce an asymmetry parameter, $\sigma$, extracted by fitting the gate-voltage dependence of the thermocurrent $I_\mathrm{th}(V_\mathrm{g})$ to the form derived in Ref.~\cite{Pyurbeevananolett.1c03591} (see SI):
\begin{equation}
    I_\mathrm{th}(V_\mathrm{g})=A \varepsilon \frac{ \dd T}{T^2}  f(\varepsilon-T \sigma) \left( 1-f(\varepsilon) \right).
\end{equation}
\noindent In the regime of weak electron-electron interactions where Onsager symmetry (see SI) applies, $\sigma$ can be interpreted as the entropy change associated with adding an electron to the system. In the Kondo regime, the situation changes: conductance develops a plateau only on one side of the CDP, signaling  breakdown of Onsager symmetry. Moreover, charge transport no longer occurs through sequential tunneling via the level at $\varepsilon$, but through higher-order co-tunneling processes. In this regime, $\sigma$ no longer reflects a thermodynamic entropy difference but instead serves as a phenomenological measure of the asymmetry in the thermoelectric signal.

However, despite its lack of an immediate physical interpretation, $\sigma$ remains an appropriate parameter for characterising the performance of the device as a heat engine, as the form of conductance and thermocurrent in Ref.~\cite{Pyurbeevananolett.1c03591} gives an adequate description of $G$ and $I_\mathrm{th}$ in the gate voltage range of non-zero thermocurrent, where the heat engine operates. The Kondo effect-induced deviation in conductance is most prominent where thermocurrent is close to zero (see Fig.~2b,c). We display the asymmetry parameter as a function of magnetic field in Fig.~\ref{fig2}f. The data indicates that $\sigma$ remains approximately constant for magnetic fields below $B \approx B_\mathrm{th}$. For $B > B_\mathrm{th}$, $\sigma$ exhibits a monotonic decrease with $B$.

In the following, we calculate the power output, $P$, and efficiency, $\eta$, of the molecular heat engine from measurements performed with different $R_\mathrm{load}$ and under different magnetic fields. While it is possible to estimate the thermal conductivity of quantum dots experimentally~\cite{Pekola_heatflowinQD}, our device design does not permit direct measurements. Therefore, we estimate the heat flow, $J_Q$, using the phenomenological Onsager model, valid for small energy level spacings (see SI). All required input parameters (tunnel couplings to reservoirs, $\Gamma_\mathrm{hot}$ and $\Gamma_\mathrm{cold}$, $\Delta T$) were obtained from fits to the experimental conductance and thermocurrent data (see SI).

\begin{figure*}[h!]
    \includegraphics[width=1.0\textwidth]{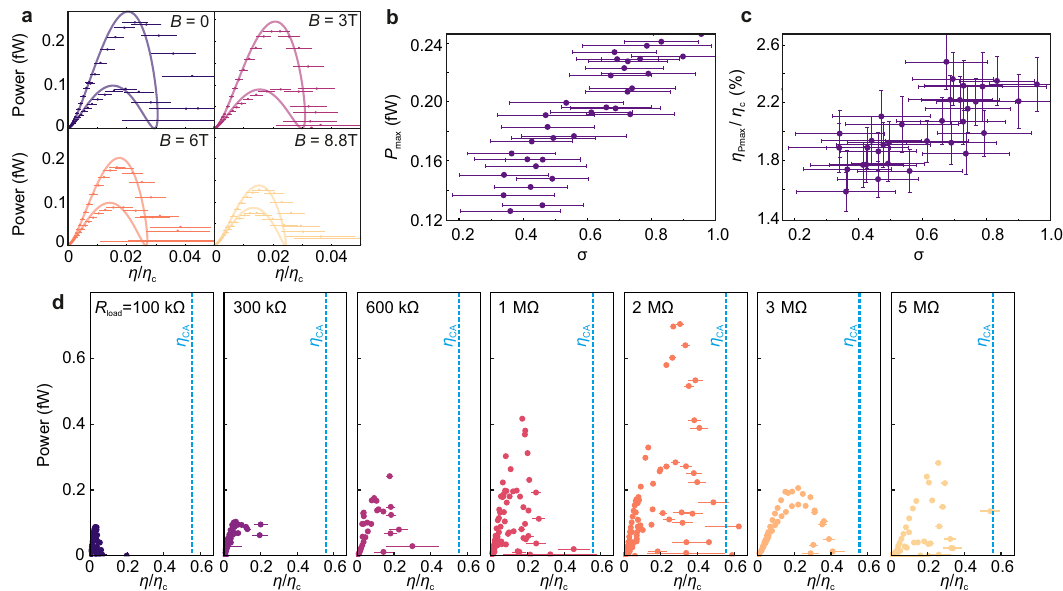}
    \caption{\label{fig:6_3Efficiency} (a) Experimental and theoretical parametric plot of power output as a function of extracted efficiency at different magnetic fields. Here, each point is taken at a different gate voltage. (b) Maximum power output and (c) efficiency at maximum power as a function of asymmetry parameter. (d) Power output as a function of efficiency for different load resistors.}
    
\end{figure*}

Figure \ref{fig:6_3Efficiency}a shows $P$ as a function of $\eta$ normalized by the Carnot efficiency $\eta_C = 1- T_\mathrm{C}/T_\mathrm{H}$ with a series resistor of \( R_\mathrm{load} = 100 \)~k\( \Omega \) recorded at different $B$-fields. Each data point in the plot corresponds to a specific configuration of the molecular level relative to the electrochemical potentials $\mu_\mathrm{h}$ and $\mu_\mathrm{c}$ of the leads. By adjusting $\varepsilon$ through changing the gate voltage, it is possible to favor either maximum thermoelectric efficiency $\eta$, or maximum output power, $P_\mathrm{max}$. From this data, we extract the efficiency at maximum power, $\eta_{P_\mathrm{max}}$, which is plotted in Figure 3b as a function of the asymmetry parameter $\sigma$. We observe that $\eta_{P_\mathrm{max}}$ reaches approximately $\text{2.3}\%$ of the Curzon–Ahlborn limit, $\eta_\mathrm{CA} = 1 - \sqrt{T_\mathrm{c}/T_\mathrm{h}}$, a benchmark for endoreversible heat engines. Importantly, $\eta_{P_\mathrm{max}}$ exhibits a clear increase with increasing $\sigma$ (decreasing magnetic field strength), indicating an impact of Kondo correlations on thermodynamic performance as will be discussed below. In addition, we find that $\eta_{P_\mathrm{max}}$ can be further optimized by tuning the external load resistance, $R_\mathrm{load}$, as shown in Fig.~\ref{fig:6_3Efficiency}d. A maximum of $\eta_{P_\mathrm{max}} \approx 0.53\, \eta_\mathrm{CA}$ is achieved for $R_\mathrm{load} = 2\,\mathrm{M}\Omega$. For larger $R_\mathrm{load}$ the efficiency decreases again.

\section{Discussion}
Our measurements show that the application of a magnetic field reduces the output power of the molecular heat engine by approximately 50\%. Magnetic field–induced changes in thermoelectric performance have previously been reported for radical molecules~\cite{Pyurbeevananolett.1c03591}, where they were attributed to variations in the spin entropy of different charge states. However, this model does not account for our observations. Based on the observed asymmetry in the thermocurrent, one would expect a higher entropy in the $N+1$ charge state than in the $N$ state. Yet, our transport data indicate a doublet-to-singlet transition upon adding an electron -- i.e., a transition from an $N$-electron doublet to an $(N+1)$-electron singlet ground state -- implying the opposite entropy trend. We therefore attribute the observed thermocurrent asymmetry -- and its magnetic field dependence -- to Kondo correlations. 

Asymmetric thermoelectric line shapes arising from spin-correlated transport have been reported in quantum dots formed in two-dimensional electron gases~\cite{Scheibner05}, semiconductor nanowires~\cite{svilans18}, and gold nanoparticle junctions~\cite{Dutta2019}. In the weak electronic coupling regime, where the magnetic field exceeds the Kondo temperature ($B \gg B_\mathrm{th}$) and where spin correlations are suppressed, the thermovoltage exhibits both positive and negative contributions near conductance peaks. This behavior arises from two distinct transport mechanisms: (1) a linear increase in thermovoltage at the center of a conductance peak due to sequential tunneling, and (2) a rapid decrease between peaks associated with cotunneling processes. In this regime, our data qualitatively follow the negative parametric derivative of the conductance, as described by Mott’s relation (see SI). In contrast, in the presence of strong spin correlations, our data—consistent with earlier studies~\cite{Scheibner05}—show clear deviations from Mott’s prediction. These deviations lead to an enhanced thermocurrent $I_\mathrm{th}$ and, consequently, an increased output power of the molecular heat engine for $B < B_\mathrm{th}$.

The enhanced thermoelectric output observed in our single-molecule heat engine in the Kondo regime likely originates from the sharp many-body resonance that forms near the Fermi level due to strong electron correlations. This resonance significantly modifies the transmission function, acting as a narrow energy filter for charge carriers. The thermoelectric response is enhanced when the resonance is slightly asymmetric with respect to the chemical potential and tuned to minimize entropy production by charge transport. Such asymmetry and tuning can arise from gate voltage control or asymmetric coupling to the electrodes, and are known to amplify the Seebeck coefficient and thermoelectric power output, as shown in theoretical studies based on the Anderson impurity model~\cite{CostiZlati,DongLei}.

We further analyze the performance of the molecular heat engine in the Kondo regime by comparing numerical simulations (see  SI), which incorporate the asymmetry parameter $\sigma$, to experimental data. The comparison reveals that increasing $\sigma$ enhances heat engine performance: it leads to higher maximum power output, improved power generation in non-optimized regimes, nearly constant efficiency at maximum power, and a slight increase in the maximum achievable efficiency.

\section{Conclusion}
We have demonstrated a single-molecule particle-exchange heat engine based on a diradical molecule operating at cryogenic temperatures, in which energy filtering is governed by a discrete molecular orbital strongly renormalized by many-body interactions. By combining electrical and thermoelectric measurements under tunable magnetic fields, we revealed that Kondo correlations – a hallmark of strong electron-electron interactions – significantly enhance the power output of the heat engine. Strikingly, we find that in the presence of Kondo correlations, the molecular engine achieves an efficiency of approximately 53\% of the Curzon–Ahlborn limit, a performance close to the best quantum dot heat engines reported to date. Our findings show that quantum many-body effects, often regarded as detrimental to coherent transport, can be harnessed to boost thermoelectric performance at the nanoscale. Looking forward, our work opens new pathways to engineer high-performance molecular heat engines by tuning spin states, electronic degeneracies, and correlation effects via chemical design and external fields.

\newpage

\section*{Author Contributions}{
  S.V. fabricated the devices, performed the electrical/thermoelectric measurements, validated and analyzed the experimental data, and supervised R.C. on fabrication of devices and performing measurements. R.C. performed the electrical/thermoelectric measurements and analyzed the experimental data. E.P. implemented a theoretical model for heat flow calculations and supported data analysis. S.F., M.V-V., and D.P. synthesized  the molecule. T.B. supported the experiment. P.G. and H.S.J.v.d.Z conceived and initiated the project. The manuscript was written through the contributions of all authors. All authors have approved the final version of the manuscript.
}
\begin{acknowledgement}
The authors acknowledge the financial support from the EU (FET-767187-QuIET, ERC-StG-10104144-MOUNTAIN, ERC Synergy Grant MolDAM (no. 951519)), from the F.R.S.-FNRS of Belgium (FNRS-CQ-1.C044.21-SMARD, FNRS-CDR-J.006823.F1-SiMolHeat), from the Netherlands Organisation for Scientific Research (NWO/OCW), as part of the Frontiers of Nanoscience program, from SPRING (EU Horizon 2020, project 863098), from Atypical (EU Pathfinder, project 101099098), from the Spanish Agencia Estatal de Investigacion (PID2022-140845OB-C62),
the Xunta de Galicia (Centro de Investigacion do Sistema Universitario de Galicia, 2023-2027, ED431G 2023/03), and the European Union (European Regional Development
FundERDF).
E.P. is grateful to the Azrieli Foundation for the award of an Azrieli Fellowship.

\end{acknowledgement}

\section*{Data Availability Statement}{
The complete data set of experiments with
description and code are available free of charge.
}
\begin{suppinfo}
The Supporting Information is available free of charge and contains: Methods section (device fabrication, molecular data, measurement protocol); DC stability diagram; characterization of Kondo peak, inelastic tunneling spectra at the right of the charge degeneracy point (CDP), evolution of charge degeneracy point  movement in a magnetic field, and thermocurrent at zero bias; Theroretical considertions for heat flow estimation; Description of data analysis and fitting protocol.

\end{suppinfo}

\section*{Note}
The authors have no conflicts to disclose.


\newpage
\section{Supplementary Materials \\
A Single-Molecule Quantum Heat Engine }

\subsection{Methods}
\subsubsection{Device}
\begin{figure*}[h!]
    \includegraphics[width=1.0\textwidth]{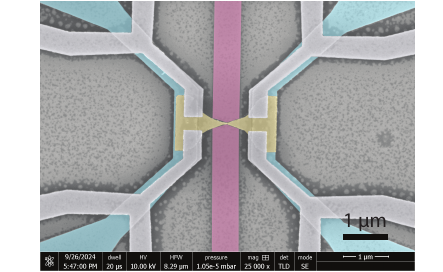}
    \caption{ \label{SEM} Scanning  electron microscope picture of the sample.}
   
\end{figure*}
For this study, we used a three-terminal electromigrated break junction (EMBJ) device with two embedded local heaters near a gold bridge~\cite{Gehring2021}. The schematic representation of the device is provided in Figure~ 1e (main text). In Fig.~\ref{SEM}, we present the scanning electron microscope (SEM) picture of the molecular device, acquired with a FEI NovaNanoSEM immediately after low temperature measurements.

The main components of the device are highlighted with color and  include the 8~nm Pd gate (pink), 27~nm Pd heaters (blue), 12~nm $\mathrm{Al}_2 \mathrm{O}_3$ protective layer (not indicated in SEM picture), 13~nm Au bridge (yellow), and 50~nm Au contacts (grey).  More details on the fabrication process are available in our prior publication~\cite {Gehring_effheating}.  The black dots around the sample were not present after sample fabrication and are attributed to molecular deposition.

We deposited the \textbf{SMe-2OS} molecule (0.1~mM solution of the molecule in dichloromethane (DCM)) on the sample using the drop-casting method. The gold bridge was electromigrated~\cite{Oneil_migration} to a resistance of 4~k$\Omega$,  and then left for self-migration (at room temperature, $10^{-2}$~mbar) for approximately 10 minutes, followed by cooling to the base temperature of 1.8~K. 

\subsubsection{Molecule}

In this work, we utilized the polycyclic aromatic hydrocarbon all-organic diradical \textbf{SMe-2OS} molecule that was synthesized by the group of Diego Pe$\mathrm{\tilde{n}}$a (see~\cite{STM_2OS} for synthesis details). The molecule consists of two fluorenyl moieties hosting a radical center each and is linked together by two anthracene units (see Fig.~1d main text). The shape of the molecule is highly non-planar, which helps to protect its radical properties and determines the interaction between the radical parts.  It also has  sulfur stabilizing groups that play the role of anchors to the gold~\cite{Frisenda_different_anchors} and help to increase the yield of the junction formation~\cite{Thomas_Baum_thesis}.  This molecule was investigated before using mechanical break junctions~\cite{Thomas_Baum_thesis} and scanning tunneling microscope~\cite{STM_2OS}. It demonstrated  Kondo resonances and inelastic electron tunneling spectroscopy (IETS) signatures (singlet-triplet excitation energy $\sim$ 3~meV). These properties were found to be sensitive to the molecular configuration within the junctions, with different configurations leading to a different coupling between spins and different observed phenomena.
\subsubsection{Thermoelectric measurements}
Measurements  were performed with an AC double lock-in technique at low temperatures. The detailed description of the measurement methodology can be found in~\cite{Gehring2021}. In the main text of the current work,  we present a schematic representation of the measurement (see Fig.~1e) and briefly describe it as following: a small AC component of 50~$\mu$V voltage bias ($\omega_1=$13~Hz)  is applied simultaneously with a current through the heater ($\omega_2=$3~Hz) for different gate and bias voltages. The signal is then demodulated with two lock-in amplifiers and a Keithley so that the differential conductance $G$ ($\omega_1$), thermocurrent, $I_\mathrm{th}$ (2$\omega_2$), and current I (DC) are simultaneously recorded. This method is valid once the temperature gradient across the junction is proportional to the power dissipated in the heater. It enables synchronous recording of transport quantities and allows extraction of the system's characteristics such as power factor, remaining insensitive to "molecular jumps". We apply a prefactor ($-2\sqrt{2}$) to $I_\mathrm{th}$ to account for the method specifics (-2 from proportionality between the heater current and the temperature gradient, and $\sqrt{2}$ related to the fact that a lock-in records root square mean values).

 All measurements were carried out in a $\mathrm{He}^4$ inset with a magnet (magnetic field up to 9~T) and a heater near the cold finger (to control temperature in the range 1.7..250~K). To quantify the performance of the molecular heat engine,  different $R_\mathrm{load}$ were connected in series, and thermocurrent was measured as a function of gate voltage and power dissipated in the heater. 
 

 \newpage
\subsection{DC stability diagram}
\begin{figure*}[h!]
    \includegraphics[width=1.0\textwidth]{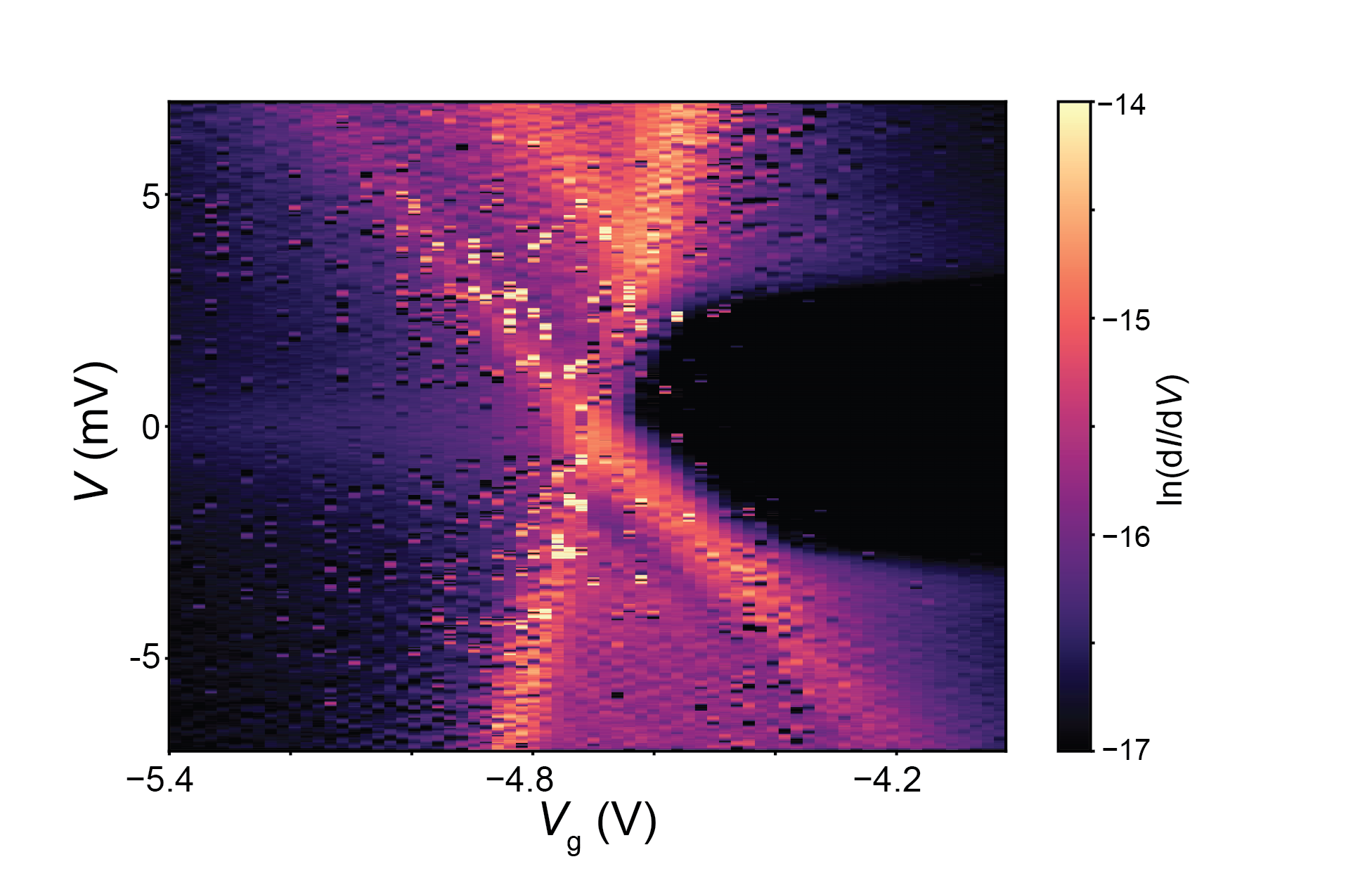}
    \caption{ \label{DC_stability} Stability diagram measured in DC to be compared to the AC measured stability diagram in the main text (Fig.~2a).}
   
\end{figure*}
In Fig.~\ref{DC_stability}, we display a DC stability diagram of our device at 1.8~K. Here, the current is first measured as a function of bias and gate voltage, then processed with the Savitzky-Golay filter to extract the differential conductance. To emphasize the physical observations -- such as Coulomb diamond edges, Kondo, excitations, and co-tunneling lines -- the color scale represents the natural logarithm of the differential conductance. The results match those displayed on Fig.~2a (main text), confirming the consistency across the AC double lock-in method and the DC measurements. Also, during this measurement, the environment was quieter, allowing us to follow the evolution of co-tunneling lines and Kondo peak next to diamond edges. 

Here, we summarize the key observations from Fig.~\ref{DC_stability} and 2a, b (main text):
\begin{itemize}
    \item CDP is around -4.7~V; there is only one CDP in the accessible gate voltage range (-7.5 .. 7.5~V).
    \item Negative diamond edge slope 0.04; positive diamond edge slope 0.012; gate coupling 0.009.
    \item Pronounced excitation parallel to negative slope at the positive bias voltages with a high conductance at around 0.5~$\mu$S; the same excitation at the negative bias voltages is faintly visible.
    \item Highly asymmetric molecule-lead coupling; total $\Gamma\approx 0.53~$meV estimated from the Coulomb peak conductance vs. gate voltage at zero bias.
    \item Thermocurrent signal in the investigated  $V$-$V_\mathrm{g}$ space ($V=-7..7$~meV, $V_\mathrm{g}=-5.5..-4.0$~V) reaches a maximum value of 300~pA.
    \item Thermocurrent changes sign on the edge of the diamond that corresponds to the lead where we apply the heater.
    \item  $I_\mathrm{th}$ is positive for IETS $V>0$ and negative for $V<0$.
\end{itemize}
\newpage

\subsection{Kondo peak}
\begin{figure*}[h!]
    \includegraphics[width=1.0\textwidth]{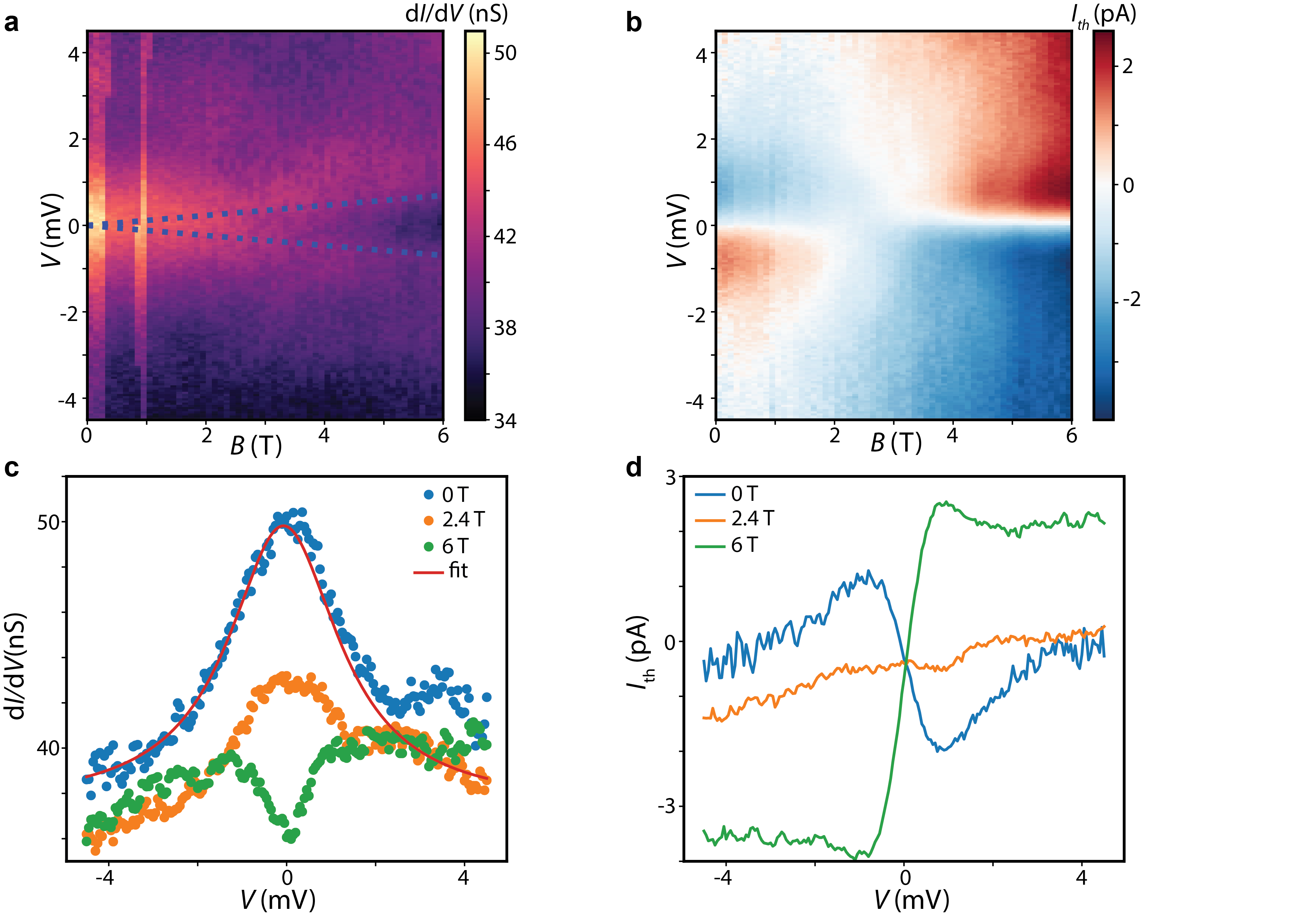}
    \caption{ \label{fig:6_6config1_Kondo} a,b. Magnetic dependence of the Kondo peak at $V_\mathrm{g}=-5.6$~V showing a Zeeman splitting in differential conductance (a) and change of the sign in thermocurrent (b).  c,. For clarity, $\mathrm{d}I/\mathrm{d}V$ (c) and $I_\mathrm{th}$ vs. voltage cuts (d) are shown for 0~T (blue), 2.4~T (orange), and 6~T (green). Thermocurrent measurements   indicate that the critical Kondo field at which splitting occures is 2.4~T. Dashed blue line   in (a) illustrates the  Zeeman effect for a spin 1/2 system with  $g=2$. Solid red line in (c) represents a Lorentzian fit to the experimental data and yields a full width half maximum (FWHM) of approximately  3~meV. }
    
\end{figure*}
To investigate the zero-bias peak for $V_\mathrm{g}<-4.7$~V, we recorded current-voltage characteristics at $V_\mathrm{g}=-5.6~$V for different magnetic fields from 0 to 6~T. In Figs.~\ref{fig:6_6config1_Kondo}a and \ref{fig:6_6config1_Kondo}b we demonstrate how the signal changes with an increase in the magnetic field: differential conductance decreases and the peak broadens with increasing magnetic field and shows the Zeeman splitting. At the highest magnetic field, the difference in energy between the ground and excited state is proportional to the magnetic field, and the distance between the peaks can be written as~\cite{Lee_oldAndersonmodel}: $\Delta E= 2 g\mu_\mathrm{B} S B$, where $g$ is the g-factor, $\mu_\mathrm{B}$ is the Bohr magneton, $S$ is the spin, and $B$ is the magnetic field. A good agreement with the measured data is found if we use $S=\frac{1}{2}$ and $g\approx 2$ (see blue dotted lines in Fig.~\ref{fig:6_6config1_Kondo}a). From the conductance measurements, it is difficult to estimate at which magnetic field the splitting starts to occur. Meanwhile, thermocurrent allows extraction of this value as the $I_\mathrm{th}$ changes its sign at the transition point~\cite{Hsu22}. As shown in Fig.~\ref{fig:6_6config1_Kondo}d, the splitting occurs around 2.4~T, providing an estimation of $T_\mathrm{K}=$ 4.3~K. Also, Kondo temperature is obtained by fitting  Lorentzian to the data (see red line in Fig.~\ref{fig:6_6config1_Kondo}c) and extracting $T_\mathrm{K}$ from~\cite{Nagaoka_Kondotemp} $\mathrm{FWHM}=2\sqrt{(\pi k_\mathrm{B}T)^2+2(k_\mathrm{B}T_\mathrm{K})^2}$ (for FWHM=3~meV $T_\mathrm{K}=$11.6~K). The difference in extracted temperatures can be attributed to a sudden shift in the position of the charge degeneracy point. Repeating the I-V at $V_\mathrm{g}$=-5.6~V and at 0~T yielded a curve with FWHM=1.5~mV, corresponding to  $T_\mathrm{K}=4.8$~K, which is in better agreement with the thermopower measurements.

\newpage
\subsection{Inelastic tunneling spectra at the right of the CDP}
\begin{figure*}[h!]
    \includegraphics[width=1.0\textwidth]{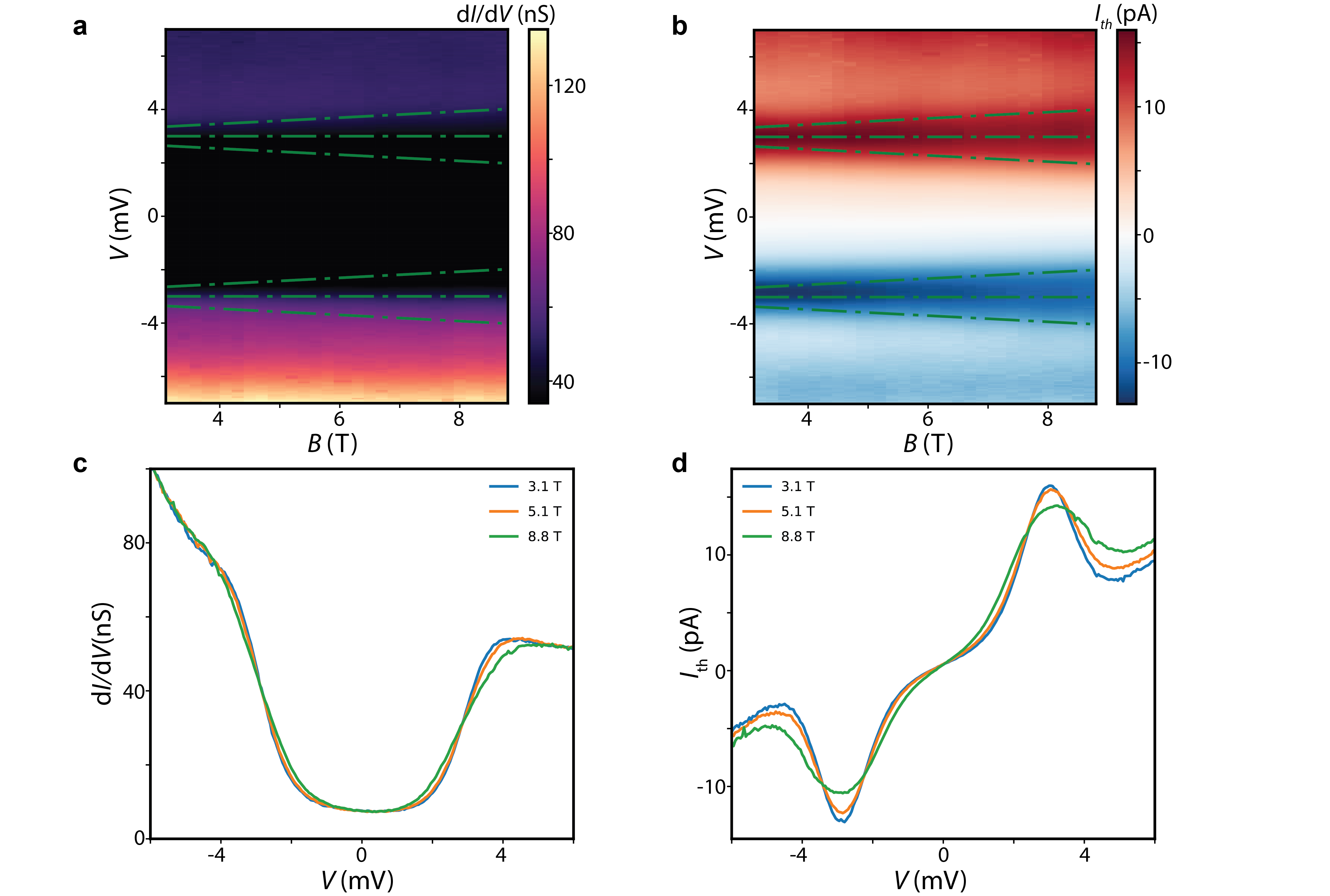}
    \caption{ \label{fig:6_7config1_IETS}a-d. Investigation of the IETS signal with increasing magnetic field at $V_\mathrm{g}=-4.15$~V. Differential conductance (a) of the peaks is broadening with the increase of the magnetic field. At the same time, the maximum/minimum peaks of the thermocurrent (b) also exhibit broadening and a decrease in amplitude with the increase of  the magnetic field. Differential conductance (c) and thermocurrent (d) cuts at 3.1~T (blue), 5.1~T (orange) and 8.8~T (green) highlight the observed evolution. Due to temperature smearing, the transitions from the ground to excited state are not distinguishable. The dotted green lines in (a,b) correspond to the Zeeman shifts (g=2) of the singlet to triplet excitations. }    
\end{figure*}
On the right-hand side of the CDP, we observe inelastic cotunneling excitations which in literature are also often referred to as inelastic electron tunneling spectroscopy (IETS) features. We recorded their behavior as a function of the magnetic field. Differential conductance and thermocurrent are illustrated in Figs.~\ref{fig:6_7config1_IETS}a-d. The IETS signal broadens but there is no clear splitting of the excitation. The broadening of the signal reaches 0.9~meV at the highest applied magnetic field of 8.8~T. At the same time, for the base temperature 3$k_\mathrm{B}T\approx0.45$~meV which would give the same broadening of around 0.9~meV. Therefore, to reveal the transitions, a lower base temperature or a higher magnetic field is required. However, considering $S=\frac{1}{2}$ (Kondo) on the left part of the CDP, the moving of the charge degeneracy point to the more positive gate voltages (see "CDP vs. B" below) indicates that the ground state is singlet ($S=0$).  The  excited state is then a triplet and the interaction between the radical parts in the ground state is antiferromagnetic.  The exchange coupling is about 3~meV (the dotted green lines in Figs.~\ref{fig:6_7config1_IETS}a,~\ref{fig:6_7config1_IETS}b correspond to the Zeeman shifts ($g=2$) of the excitations). 
\newpage
\subsection{Evolution of the CDP in a magnetic field}
\begin{figure*}[h!]
    \includegraphics[width=1.0\textwidth]{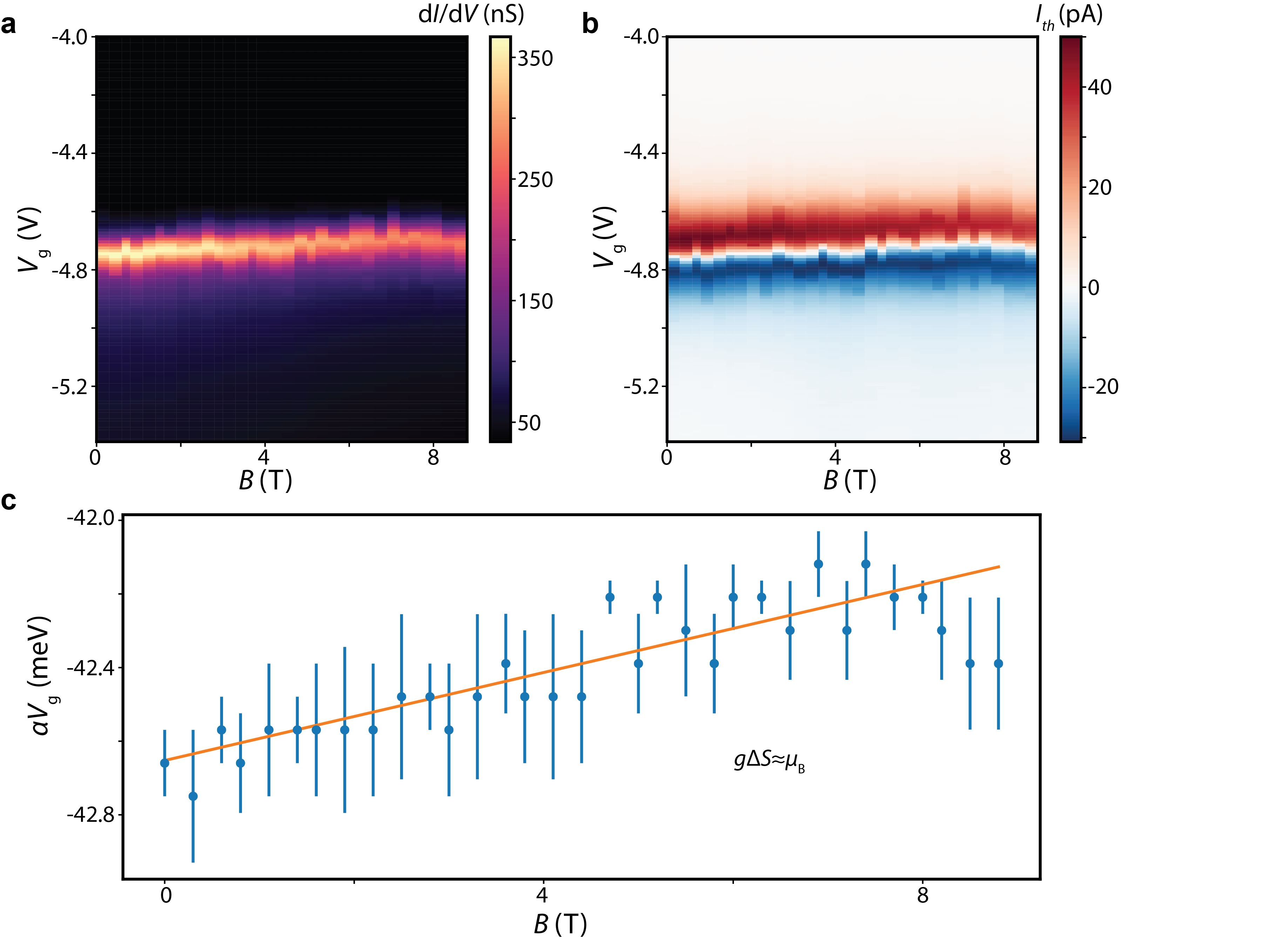}
    \caption{\label{fig:6_8config1_CDP}a,b.Colormaps of differential conductance (a) and thermocurrent (b) for  different magnetic fields and gate voltages at zero bias voltage. c. Magnetic field dependence of CDP position where the vertical axis has been converted into an energy axis by multiplication the gate voltage with the gate coupling parameter ($\alpha=0.009$). The orange line represents a linear fit to the data, yielding $g\Delta S\approx1$.}
    
\end{figure*}
We performed zero-bias gate traces for different magnetic fields up to 8.8~T. Differential conductance and thermocurrent as a function of gate voltage for the different magnetic fields are measured simultaneously and the results are  depicted in Figs.~\ref{fig:6_8config1_CDP}a and \ref{fig:6_8config1_CDP}b, respectively. Once the energy level of the molecule is in resonance with the electrochemical potential of the leads, a pronounced peak in the differential conductance appears.  This Coulomb peak decreases in amplitude and moves to less negative gate values with increasing magnetic field (Fig.~\ref{fig:6_8config1_CDP}a). The complete evolution of the maximum of the peak is shown on Fig.~\ref{fig:6_8config1_CDP}c. Here, the energetic displacement of CDP is recalculated from the gate voltage and the gate coupling parameter, $\alpha$, which we extract from the edges of a Coulomb diamond. During the gate sweeps, the junction environment experiences fluctuating charging effects that lead to an uncertainty in its position (see error bars in Fig.~\ref{fig:6_8config1_CDP}c). Furthermore, we fit the equation $\alpha \Delta(\mathrm{e}V_\mathrm{g})=\mu_\mathrm{B}(g^NS^N-g^{N-1}S^{N-1})B$ to the experimental data. To avoid influence of Kondo correlations and eliminate "charge-jump" discontinuities in the data, we restrict our analysis to the magnetic field range from 2.4~T to 8.2~T. Our analysis is consistent with $g\approx2$ and $\Delta S=\frac{1}{2}$ and shows that the region to the right of the CDP ($N+1$) corresponds to a singlet ground state, the $N$ state is doublet. 

As illustrated in Fig.~\ref{fig:6_8config1_CDP}b, the thermocurrent vanishes deep in the Coulomb blockaded region and reaches extrema near the CDP, where it changes sign at resonance. Different carriers are responsible for the thermocurrent at $N$ ($I_\mathrm{th}<0$, electrons) and $N+1$ ($I_\mathrm{th}>0$, holes) regions. We observe an asymmetry between amplitudes in both regions that we relate to an entropy change.  We further investigate the role of magnetic field and Kondo correlations to the thermocurrent and performance of the molecular heat engine in the following sections. 

\newpage
\subsection{Thermocurrent}
\begin{figure*}[h!]
    \includegraphics[width=1.0\textwidth]{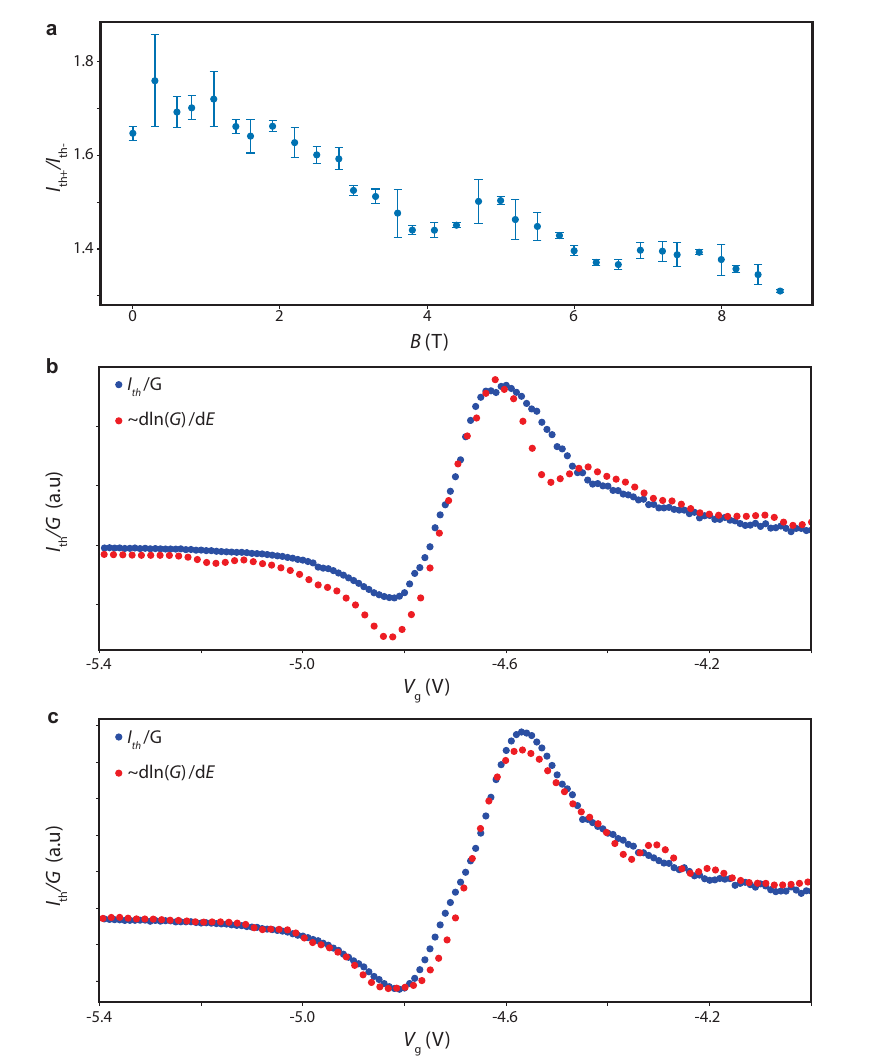}
    \caption{\label{fig:SI6_thermopower}a. Ratio between maximum positive and negative thermocurrent at different magnetic fields. b,c.  Negative thermovoltage as a function of gate voltage measured with the double lock-in technique, $\frac{I_\mathrm{th}}{G}$ (blue points) and evaluated from Mott formula (red points) at 0~T (b) and 8.8~T (c).}
    
\end{figure*}

In Fig.~\ref{fig:SI6_thermopower}a we display how the ratio between the maximum positive, $I_\mathrm{th+}$, and negative, $I_\mathrm{th-}$, values of the thermocurrent evolves with the magnetic field.  These values are extracted from the gate voltage sweeps at zero bias, where $I_\mathrm{th+}$ and $I_\mathrm{th-}$ represent the peak thermocurrent contributions from holes and electrons, respectively. We estimate the error as the difference between the readings obtained during sweeps back and forth.   The ratio stays more or less flat in the magnetic field range below the critical Kondo field (2.4~T) and then tends to drop to unity with increasing magnetic field (reaching 1.3 at 8.8~T). There are also oscillations that can be observed in the figure around 4~T and 6.6~T. Their exact origin is not  understood.

Subsequently, we focus our attention on the thermopower coefficient $S$ and evaluate to what extent it can be described by the Mott relation. For this purpose, we plot $\frac{I_\mathrm{th}}{G}$ as a function of gate voltage (see blue dots in Fig.~\ref{fig:SI6_thermopower}b,c).  The Mott relation $S\sim \frac{\mathrm{d}G(E)}{\mathrm{d}E}$ is plotted as red dots in Fig.~\ref{fig:SI6_thermopower}b,c. At high magnetic fields,there is a good correspondence between the curves (Fig.~\ref{fig:SI6_thermopower}c), whereas in the absence of a magnetic field, the Mott formula does not reproduce the measured thermo $\frac{I_\mathrm{th}}{G}$ well (Fig.~\ref{fig:SI6_thermopower}b).

\newpage
\subsection{Theory}

\subsubsection{Heat engine operating parameters}
For the theoretical analysis, we consider the molecular heat engine to operate in the linear regime, where the device operation is fully determined by conductance and thermoelectric susceptibility, $G$ and $L$ (the $L_{00}$ and $L_{01}$ coefficients of the Onsager matrix), which are properties of the microscopic dynamics of the device; and $\dd T$ and $\dd V$, both much smaller than $\kb T$ -- the operating parameters. 
$\dd T$ is free and externally imposed, while $\dd V$ is determined to satisfy the Ohm's law for the load resistor.  

Since high heat engine performance requires a narrow energy window for particle exchange, we can assume that transmission is characterised by a single non-broadened energy level at $\varepsilon$. Then, from the symmetry of the Onsager matrix: 
\begin{equation}
\label{eq-Onsager-crit}
    L=\frac{\varepsilon}{T}G.
\end{equation}
Using the above, we determine the power and efficiency of a heat engine under our above assumptions. The current through the resistor is equal to:
\begin{equation}
    I=L \dd T - G \dd V =L \dd T - G R I,
\end{equation}
from which we find the following self-consistent expression:
\begin{equation}
    I=\frac{L \dd T}{1+G R}.
\end{equation}
This allows us to fine-tune the energetic parameters -- power:
\begin{equation}
\label{eq-power}
    P=I^2 R=\frac{\dd T^2}{T^2} \frac{\varepsilon^2 G^2 R}{(1+G R) ^2},
\end{equation}
and efficiency:
\begin{equation}
    \eta=\frac{P}{\dot{Q_H}}\approx \frac{I^2 R}{I \varepsilon}=\frac{IR}{\varepsilon}=\frac{\dd T}{T} \frac{GR}{(1+GR)},
\end{equation}
which, relative to the Carnot efficiency (in the limit of small $\dd T$) can be written as:
\begin{equation}
\label{eq-eff}
    \frac{\eta}{\eta_C}=\frac{GR}{(1+GR)}
\end{equation}
This result (Eq. \ref{eq-eff}) has as intuitive interpretation: since the model does not include the passive heat flow between the baths, and $G R \rightarrow \infty$, the load is large and the heat engine is operated slowly, so that $\eta/ \eta_C \rightarrow 1$. For small $GR$, on the other hand, $\eta/ \eta_C \rightarrow 0$.

\subsubsection{Device description}
In our previous work \cite{Pyurbeevananolett.1c03591} we have demonstrated that for a range of quantum dots or molecular SETs with two energetically accessible charge states in the sequential tunneling regime, conductance and thermoelectic susceptibility can be expressed as:
\begin{equation}
\label{eq-GS}
    G=\frac{A}{T} f(\varepsilon-T \Delta S) \left( 1-f(\varepsilon) \right),
\end{equation}
\begin{equation}
\label{eq-LS}
    L=\frac{\varepsilon}{T^2} A f(\varepsilon-T \Delta S) \left( 1-f(\varepsilon) \right),
\end{equation}
where $A$ is a function of coupling strengths, $\varepsilon$ is the difference between the transition energy and the equilibrium chemical potential of the electrodes, $f$ is the Fermi distribution, and $\Delta S$ is the entropy difference between the charge states.  

The conditions for Eqs.~8, 9 
to hold include: (i) the dynamics to be solely described by transitions between fixed sets of energy levels; (ii) the energy level spacing in each of the charge states to be much smaller than the $\kb T$; and (iii) the absence of selection rules for the transitions or relaxation within a single charge state. 

However, the utility of the above form of $G$ and $L$ can be much broader if the entropic interpretation is to be set aside. The thermocurrent, $L \dd T$ passes through zero at $\varepsilon=0$, changing sign, and approaches zero again for $\varepsilon\gg k_\mathrm{b} T$. This, together with  Eq.\ref{eq-Onsager-crit}, a property of the Onsager matrix, implies that similar expressions:
\begin{equation}
\label{eq-G}
    G=\frac{A}{T} f(\varepsilon-T \sigma) \left( 1-f(\varepsilon) \right),
\end{equation}
\begin{equation}
\label{eq-L}
    L=\frac{\varepsilon}{T^2} A f(\varepsilon-T \sigma) \left( 1-f(\varepsilon) \right),
\end{equation}
hold for a more general case of sequential tunnelling current through a single level, if $\sigma$ is no longer the entropy difference, but simply a fitting parameter determining the degree of asymmetry of the thermocurrent trace.

Note, that the Onsager symmetry (Eq. \ref{eq-Onsager-crit}) does not hold in our case, since the Kondo effect falls outside of the assumption required for it, as the charge in the Kondo regime is not transmitted classically through the energy level at $\varepsilon$, and in practice presents with a conductance plateau on one side of the charge degeneracy point, which is not reflected in the thermocurrent. However, as the thermocurrent, and thus efficiency and generated power far from the charge degeneracy point is zero, the difference created by non-zero conductance at zero thermocurrent is negligible, and Eqs.~\ref{eq-G},\ref{eq-L} can still be used to describe our molecular system for the purposes of characterising it as a heat engine.

\subsubsection{The effect of the asymmetry parameter on heat engine operation}
\begin{figure}[h]
    \centering
    \includegraphics[width=0.8\textwidth]{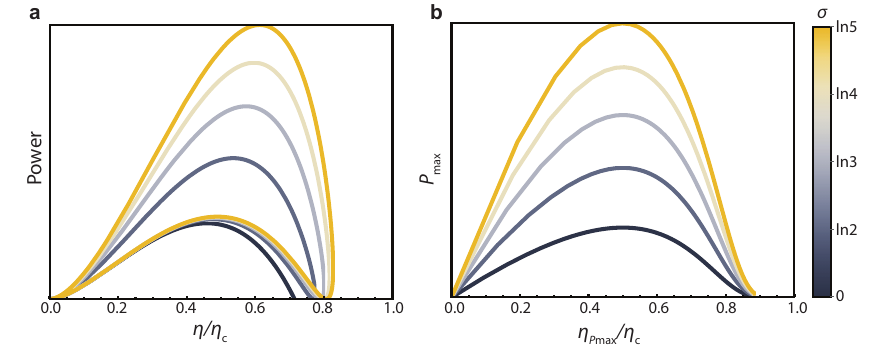}
    \caption{Theoretical results for the quantum dot heat engine properties as a function of the asymmetry parameter $\sigma$. The loops in the power-vs-efficiency figure consist of two separate dependencies for the positive and negative values of $\varepsilon$ spanning from $\eta/\eta_C=0$ to the maximum value.}
  \label{sigma model}
\end{figure}
Using the asymmetry-based description of the device, we can now analyse the energetic properties of a heat engine based on it and the influence of the asymmetry parameter, $\sigma$, on them. We do it numerically, by substituting Eq.~\ref{eq-G} into the expression for efficiency (Eq.~\ref{eq-eff}) and power (Eq. \ref{eq-power}) and then finding efficiency at maximum power. 

Fig~\ref{sigma model} shows the results for a single value of $R$ as an example. It can be seen that increasing $\sigma$ leads to overall improved heat engine operation: higher maximum power output, as well to higher power at non-optimised regimes, and to slightly increased maximum efficiency.   
\newpage
\subsection{Data analysis and fitting}
Here, we outline the step-by-step process for the data analysis employed in this work.
\begin{enumerate}
    \item The lever arm, $\alpha$, where $\varepsilon=\alpha (V_G -V_0)$, and $V_0$ corresponds to the position of the chemical potential of the electrodes at equilibrium, was found from the stability diagram, by comparing the detuning effects of the gate and bias voltages.
    \item In order to find the quasi-static load-independent characteristics of the device (as if conductance and thermocurrent had been measured at zero load), we correct the data by $1/(1-G(V_g)R_{\text{load}})$ in case of conductance and $(1+G(V_g)R_{\text{load}})$ in case of thermocurrent.   
    \item Each of the corrected thermocurrent traces was fitted to:
    \begin{equation}
    \label{eq-Lfit}
        I_{th}(V_g)=a (V_g-b) f\left(\frac{V_g-b}{t}-\sigma\right) \left(1-f\left(\frac{V_g-b}{t}\right) \right),
    \end{equation}
    where $V_g$ is the experimental gate voltage in the trace, and $a$, $b$, $t$ and $\sigma$ are numerical fitting parameters: $a$ corresponds to the overall amplitude of the trace, $b$ to the zero thermocurrent point ($\varepsilon=0$), $t$ to the width of the trace (note: it is not equal to the temperature), and $\sigma$ to the asymmetry of the trace.    
    \item Similarly, all conductance traces were fitted as:
    \begin{equation}
    \label{eq-Gfit}
        G(V_g)=a_G f\left(\frac{V_g-b_G}{t_G}-\sigma \right) \left(1-f\left(\frac{V_g-b_G}{t_G}\right) \right).
    \end{equation}
    Note, that the functions $f(\varepsilon-\sigma)(1-f(\varepsilon))$ and $f(\varepsilon)(1-f(\varepsilon))$ have the same functional shape of the peak and differ only in the amplitude and peak position. Therefore, in the fit, $\sigma$ would have been an excessive parameter. To avoid this issue, we used the value of $\sigma$ determined from the corresponding thermocurrent fit.   
    \item To prove the validity of the fits, $t$ and $t_G$ can be compared to each other in each trace, as well as to $\kb T/\alpha$, where $T$ is the temperature of the cryostat -- the correct value for the width parameter of the trace. Alternatively, the lever arm $\alpha$ can be found from this data, if the ambient temperature is known. 
    \item Comparing Eqs. \ref{eq-Lfit},\ref{eq-Gfit} to the expressions for $G$ and $L$ (Eqs. \ref{eq-G},\ref{eq-L}), we find that for every pair of conductance and thermocurrent traces, 
    \begin{equation}
        \frac{a}{a_G}=\frac{\alpha \Delta T}{T}.
    \end{equation}
    This allows us to find the expected Carnot efficiencies $\Delta T/T$, assuming that $\Delta T$ is small compared to the ambient temperature $T$. 

    \item We find the power output as $P=R_{\text{load}}I_m^2$, where $I_m$ is the measured (uncorrected) thermocurrent, and the heat current is $\dot{Q}=\alpha (V_g-b) I_m$, where $b$ was found from the fit of the corrected thermocurrent ($\alpha (V_g-b)=\varepsilon$). 
    \item Finally, $P/\dot{Q}$ allows us to find the efficiency of the heat engine. The Carnot efficiency has been calculated previously, allowing us to find $\eta/\eta_c$.
\end{enumerate}

Since for our data, due to the Kondo effect, the experimental conductance differed from the fitting equation (Eq. \ref{eq-Gfit}), for fitting purposes we only used the data that fell within the thermal broadening range left of the conductance peak. However, this approach can lead to systematic errors, which we see in Fig.2 in the main manuscript. It can also be seen that this error decreases with magnetic field, as expected with the Kondo effect.  


\clearpage
\bibliography{achemso-demo}

\providecommand{\noopsort}[1]{}\providecommand{\singleletter}[1]{#1}%
\providecommand{\latin}[1]{#1}
\makeatletter
\providecommand{\doi}
  {\begingroup\let\do\@makeother\dospecials
  \catcode`\{=1 \catcode`\}=2 \doi@aux}
\providecommand{\doi@aux}[1]{\endgroup\texttt{#1}}
\makeatother
\providecommand*\mcitethebibliography{\thebibliography}
\csname @ifundefined\endcsname{endmcitethebibliography}  {\let\endmcitethebibliography\endthebibliography}{}
\begin{mcitethebibliography}{34}
\providecommand*\natexlab[1]{#1}
\providecommand*\mciteSetBstSublistMode[1]{}
\providecommand*\mciteSetBstMaxWidthForm[2]{}
\providecommand*\mciteBstWouldAddEndPuncttrue
  {\def\EndOfBibitem{\unskip.}}
\providecommand*\mciteBstWouldAddEndPunctfalse
  {\let\EndOfBibitem\relax}
\providecommand*\mciteSetBstMidEndSepPunct[3]{}
\providecommand*\mciteSetBstSublistLabelBeginEnd[3]{}
\providecommand*\EndOfBibitem{}
\mciteSetBstSublistMode{f}
\mciteSetBstMaxWidthForm{subitem}{(\alph{mcitesubitemcount})}
\mciteSetBstSublistLabelBeginEnd
  {\mcitemaxwidthsubitemform\space}
  {\relax}
  {\relax}

\bibitem[Whalen \latin{et~al.}(2003)Whalen, Thompson, Bahr, Richards, and Richards]{Wha03}
Whalen,~S.; Thompson,~M.; Bahr,~D.; Richards,~C.; Richards,~R. Design, fabrication and testing of the P3 micro heat engine. \emph{Sensors and Actuators A: Physical} \textbf{2003}, \emph{104}, 290--298\relax
\mciteBstWouldAddEndPuncttrue
\mciteSetBstMidEndSepPunct{\mcitedefaultmidpunct}
{\mcitedefaultendpunct}{\mcitedefaultseppunct}\relax
\EndOfBibitem
\bibitem[Steeneken \latin{et~al.}(2011)Steeneken, Le~Phan, Goossens, Koops, Brom, van~der Avoort, and van Beek]{Ste11}
Steeneken,~P.~G.; Le~Phan,~K.; Goossens,~M.~J.; Koops,~G. E.~J.; Brom,~G. J. A.~M.; van~der Avoort,~C.; van Beek,~J. T.~M. Piezoresistive heat engine and refrigerator. \emph{Nat. Phys.} \textbf{2011}, \emph{7}, 354--359\relax
\mciteBstWouldAddEndPuncttrue
\mciteSetBstMidEndSepPunct{\mcitedefaultmidpunct}
{\mcitedefaultendpunct}{\mcitedefaultseppunct}\relax
\EndOfBibitem
\bibitem[Roßnagel \latin{et~al.}(2016)Roßnagel, Dawkins, Tolazzi, Abah, Lutz, Schmidt-Kaler, and Singer]{Ros16}
Roßnagel,~J.; Dawkins,~S.~T.; Tolazzi,~K.~N.; Abah,~O.; Lutz,~E.; Schmidt-Kaler,~F.; Singer,~K. A single-atom heat engine. \emph{Science} \textbf{2016}, \emph{352}, 325--329\relax
\mciteBstWouldAddEndPuncttrue
\mciteSetBstMidEndSepPunct{\mcitedefaultmidpunct}
{\mcitedefaultendpunct}{\mcitedefaultseppunct}\relax
\EndOfBibitem
\bibitem[Quan(2009)]{Qua09}
Quan,~H.~T. Quantum thermodynamic cycles and quantum heat engines. II. \emph{Phys. Rev. E} \textbf{2009}, \emph{79}, 041129\relax
\mciteBstWouldAddEndPuncttrue
\mciteSetBstMidEndSepPunct{\mcitedefaultmidpunct}
{\mcitedefaultendpunct}{\mcitedefaultseppunct}\relax
\EndOfBibitem
\bibitem[Feldmann and Kosloff(2003)Feldmann, and Kosloff]{Fel03}
Feldmann,~T.; Kosloff,~R. Quantum four-stroke heat engine: Thermodynamic observables in a model with intrinsic friction. \emph{Phys. Rev. E} \textbf{2003}, \emph{68}, 016101\relax
\mciteBstWouldAddEndPuncttrue
\mciteSetBstMidEndSepPunct{\mcitedefaultmidpunct}
{\mcitedefaultendpunct}{\mcitedefaultseppunct}\relax
\EndOfBibitem
\bibitem[Rezek and Kosloff(2006)Rezek, and Kosloff]{Rez06}
Rezek,~Y.; Kosloff,~R. Irreversible performance of a quantum harmonic heat engine. \emph{New Journal of Physics} \textbf{2006}, \emph{8}, 83\relax
\mciteBstWouldAddEndPuncttrue
\mciteSetBstMidEndSepPunct{\mcitedefaultmidpunct}
{\mcitedefaultendpunct}{\mcitedefaultseppunct}\relax
\EndOfBibitem
\bibitem[Mahan and Sofo(1996)Mahan, and Sofo]{Mahan96}
Mahan,~G.~D.; Sofo,~J.~O. The best thermoelectric. \emph{Proceedings of the National Academy of Sciences} \textbf{1996}, \emph{93}, 7436--7439\relax
\mciteBstWouldAddEndPuncttrue
\mciteSetBstMidEndSepPunct{\mcitedefaultmidpunct}
{\mcitedefaultendpunct}{\mcitedefaultseppunct}\relax
\EndOfBibitem
\bibitem[Josefsson \latin{et~al.}(2018)Josefsson, Svilans, Burke, Hoffmann, Fahlvik, Thelander, Leijnse, and Linke]{Josefsson2018}
Josefsson,~M.; Svilans,~A.; Burke,~A.~M.; Hoffmann,~E.~A.; Fahlvik,~S.; Thelander,~C.; Leijnse,~M.; Linke,~H. A quantum-dot heat engine operating close to the thermodynamic efficiency limits. \emph{Nature Nanotechnology} \textbf{2018}, \emph{13}, 920--924\relax
\mciteBstWouldAddEndPuncttrue
\mciteSetBstMidEndSepPunct{\mcitedefaultmidpunct}
{\mcitedefaultendpunct}{\mcitedefaultseppunct}\relax
\EndOfBibitem
\bibitem[Josefsson \latin{et~al.}(2019)Josefsson, Svilans, Linke, and Leijnse]{Josefsson2019}
Josefsson,~M.; Svilans,~A.; Linke,~H.; Leijnse,~M. Optimal power and efficiency of single quantum dot heat engines: Theory and experiment. \emph{Phys. Rev. B} \textbf{2019}, \emph{99}, 235432\relax
\mciteBstWouldAddEndPuncttrue
\mciteSetBstMidEndSepPunct{\mcitedefaultmidpunct}
{\mcitedefaultendpunct}{\mcitedefaultseppunct}\relax
\EndOfBibitem
\bibitem[Verma and Singh(2023)Verma, and Singh]{Sachin}
Verma,~S.; Singh,~A. A Strongly Correlated Quantum Dot Heat Engine with Optimal Performance: A Nonequilibrium Green's Function Approach. \emph{physica status solidi (b)} \textbf{2023}, \emph{260}, 2200608\relax
\mciteBstWouldAddEndPuncttrue
\mciteSetBstMidEndSepPunct{\mcitedefaultmidpunct}
{\mcitedefaultendpunct}{\mcitedefaultseppunct}\relax
\EndOfBibitem
\bibitem[Pyurbeeva \latin{et~al.}(2021)Pyurbeeva, Hsu, Vogel, Wegeberg, Mayor, van~der Zant, Mol, and Gehring]{Pyurbeevananolett.1c03591}
Pyurbeeva,~E.; Hsu,~C.; Vogel,~D.; Wegeberg,~C.; Mayor,~M.; van~der Zant,~H.; Mol,~J.~A.; Gehring,~P. Controlling the Entropy of a Single-Molecule Junction. \emph{Nano Letters} \textbf{2021}, \emph{21}, 9715--9719, PMID: 34766782\relax
\mciteBstWouldAddEndPuncttrue
\mciteSetBstMidEndSepPunct{\mcitedefaultmidpunct}
{\mcitedefaultendpunct}{\mcitedefaultseppunct}\relax
\EndOfBibitem
\bibitem[Pyurbeeva \latin{et~al.}(2022)Pyurbeeva, Mol, and Gehring]{Pyurbeeva22}
Pyurbeeva,~E.; Mol,~J.~A.; Gehring,~P. {Electronic measurements of entropy in meso- and nanoscale systems}. \emph{Chemical Physics Reviews} \textbf{2022}, \emph{3}, 041308\relax
\mciteBstWouldAddEndPuncttrue
\mciteSetBstMidEndSepPunct{\mcitedefaultmidpunct}
{\mcitedefaultendpunct}{\mcitedefaultseppunct}\relax
\EndOfBibitem
\bibitem[Svilans \latin{et~al.}(2018)Svilans, Josefsson, Burke, Fahlvik, Thelander, Linke, and Leijnse]{svilans18}
Svilans,~A.; Josefsson,~M.; Burke,~A.~M.; Fahlvik,~S.; Thelander,~C.; Linke,~H.; Leijnse,~M. Thermoelectric Characterization of the Kondo Resonance in Nanowire Quantum Dots. \emph{Phys. Rev. Lett.} \textbf{2018}, \emph{121}, 206801\relax
\mciteBstWouldAddEndPuncttrue
\mciteSetBstMidEndSepPunct{\mcitedefaultmidpunct}
{\mcitedefaultendpunct}{\mcitedefaultseppunct}\relax
\EndOfBibitem
\bibitem[Hsu \latin{et~al.}(2022)Hsu, Costi, Vogel, Wegeberg, Mayor, van~der Zant, and Gehring]{Hsu22}
Hsu,~C.; Costi,~T.~A.; Vogel,~D.; Wegeberg,~C.; Mayor,~M.; van~der Zant,~H. S.~J.; Gehring,~P. Magnetic-Field Universality of the Kondo Effect Revealed by Thermocurrent Spectroscopy. \emph{Phys. Rev. Lett.} \textbf{2022}, \emph{128}, 147701\relax
\mciteBstWouldAddEndPuncttrue
\mciteSetBstMidEndSepPunct{\mcitedefaultmidpunct}
{\mcitedefaultendpunct}{\mcitedefaultseppunct}\relax
\EndOfBibitem
\bibitem[Volosheniuk \latin{et~al.}(2025)Volosheniuk, Bouwmeester, Vogel, Wegeberg, Hsu, Mayor, van~der Zant, and Gehring]{Volosheniuk2025}
Volosheniuk,~S.; Bouwmeester,~D.; Vogel,~D.; Wegeberg,~C.; Hsu,~C.; Mayor,~M.; van~der Zant,~H. S.~J.; Gehring,~P. Enhancing thermoelectric output in a molecular heat engine utilizing Yu-Shiba-Rusinov bound states. \emph{Nature Communications} \textbf{2025}, \emph{16}, 3279\relax
\mciteBstWouldAddEndPuncttrue
\mciteSetBstMidEndSepPunct{\mcitedefaultmidpunct}
{\mcitedefaultendpunct}{\mcitedefaultseppunct}\relax
\EndOfBibitem
\bibitem[Peng \latin{et~al.}(2025)Peng, Chen, Xie, Ma, Lü, and Li]{D5TA01503K}
Peng,~W.; Chen,~N.; Xie,~Y.; Ma,~L.; Lü,~J.-T.; Li,~Y. Heteroatom engineering for enhancing the thermoelectric power factor of molecular junctions. \emph{J. Mater. Chem. A} \textbf{2025}, \emph{13}, 15222--15231\relax
\mciteBstWouldAddEndPuncttrue
\mciteSetBstMidEndSepPunct{\mcitedefaultmidpunct}
{\mcitedefaultendpunct}{\mcitedefaultseppunct}\relax
\EndOfBibitem
\bibitem[Ma \latin{et~al.}(2020)Ma, Nian, and L\"u]{PhysRevB.101.045410}
Ma,~L.; Nian,~L.-L.; L\"u,~J.-T. Design and optimization of a heat engine based on a porphyrin single-molecule junction with graphene electrodes. \emph{Phys. Rev. B} \textbf{2020}, \emph{101}, 045410\relax
\mciteBstWouldAddEndPuncttrue
\mciteSetBstMidEndSepPunct{\mcitedefaultmidpunct}
{\mcitedefaultendpunct}{\mcitedefaultseppunct}\relax
\EndOfBibitem
\bibitem[Gehring \latin{et~al.}(2019)Gehring, Thijssen, and van~der Zant]{Gehring2019_theory}
Gehring,~P.; Thijssen,~J.~M.; van~der Zant,~H. S.~J. Single-molecule quantum-transport phenomena in break junctions. \emph{Nature Reviews Physics} \textbf{2019}, \emph{1}, 381--396\relax
\mciteBstWouldAddEndPuncttrue
\mciteSetBstMidEndSepPunct{\mcitedefaultmidpunct}
{\mcitedefaultendpunct}{\mcitedefaultseppunct}\relax
\EndOfBibitem
\bibitem[Pyurbeeva and Mol(2021)Pyurbeeva, and Mol]{Pyurbeeva2020b}
Pyurbeeva,~E.; Mol,~J.~A. A Thermodynamic Approach to Measuring Entropy in a Few-Electron Nanodevice. \emph{Entropy} \textbf{2021}, \emph{23}\relax
\mciteBstWouldAddEndPuncttrue
\mciteSetBstMidEndSepPunct{\mcitedefaultmidpunct}
{\mcitedefaultendpunct}{\mcitedefaultseppunct}\relax
\EndOfBibitem
\bibitem[Vegliante \latin{et~al.}(2024)Vegliante, Fernández, Ortiz, Vilas-Varela, Baum, Friedrich, Romero-Lara, Aguirre, Vaxevani, Wang, Garcia~Fernandez, van~der Zant, Frederiksen, Pe{\~n}a, and Pascual]{STM_2OS}
Vegliante,~A.; Fernández,~S.; Ortiz,~R.; Vilas-Varela,~M.; Baum,~T.~Y.; Friedrich,~N.; Romero-Lara,~F.; Aguirre,~A.; Vaxevani,~K.; Wang,~D.; Garcia~Fernandez,~C.; van~der Zant,~H. S.~J.; Frederiksen,~T.; Pe{\~n}a,~D.; Pascual,~J.~I. Tuning the Spin Interaction in Nonplanar Organic Diradicals through Mechanical Manipulation. \emph{ACS Nano} \textbf{2024}, \emph{18}, 26514--26521\relax
\mciteBstWouldAddEndPuncttrue
\mciteSetBstMidEndSepPunct{\mcitedefaultmidpunct}
{\mcitedefaultendpunct}{\mcitedefaultseppunct}\relax
\EndOfBibitem
\bibitem[O’Neill \latin{et~al.}(2007)O’Neill, Osorio, and van~der Zant]{Oneil_migration}
O’Neill,~K.; Osorio,~E.~A.; van~der Zant,~H. S.~J. {Self-breaking in planar few-atom Au constrictions for nanometer-spaced electrodes}. \emph{Applied Physics Letters} \textbf{2007}, \emph{90}, 133109\relax
\mciteBstWouldAddEndPuncttrue
\mciteSetBstMidEndSepPunct{\mcitedefaultmidpunct}
{\mcitedefaultendpunct}{\mcitedefaultseppunct}\relax
\EndOfBibitem
\bibitem[Gehring \latin{et~al.}(2021)Gehring, Sowa, Hsu, de~Bruijckere, van~der Star, Le~Roy, Bogani, Gauger, and van~der Zant]{Gehring2021}
Gehring,~P.; Sowa,~J.~K.; Hsu,~C.; de~Bruijckere,~J.; van~der Star,~M.; Le~Roy,~J.~J.; Bogani,~L.; Gauger,~E.~M.; van~der Zant,~H. S.~J. Complete mapping of the thermoelectric properties of a single molecule. \emph{Nature Nanotechnology} \textbf{2021}, \emph{16}, 426--430\relax
\mciteBstWouldAddEndPuncttrue
\mciteSetBstMidEndSepPunct{\mcitedefaultmidpunct}
{\mcitedefaultendpunct}{\mcitedefaultseppunct}\relax
\EndOfBibitem
\bibitem[Volosheniuk \latin{et~al.}(2023)Volosheniuk, Bouwmeester, Hsu, van~der Zant, and Gehring]{Volo23}
Volosheniuk,~S.; Bouwmeester,~D.; Hsu,~C.; van~der Zant,~H. S.~J.; Gehring,~P. {Implementation of SNS thermometers into molecular devices for cryogenic thermoelectric experiments}. \emph{Applied Physics Letters} \textbf{2023}, \emph{122}, 103501\relax
\mciteBstWouldAddEndPuncttrue
\mciteSetBstMidEndSepPunct{\mcitedefaultmidpunct}
{\mcitedefaultendpunct}{\mcitedefaultseppunct}\relax
\EndOfBibitem
\bibitem[Dutta \latin{et~al.}(2017)Dutta, Peltonen, Antonenko, Meschke, Skvortsov, Kubala, K\"onig, Winkelmann, Courtois, and Pekola]{Pekola_heatflowinQD}
Dutta,~B.; Peltonen,~J.~T.; Antonenko,~D.~S.; Meschke,~M.; Skvortsov,~M.~A.; Kubala,~B.; K\"onig,~J.; Winkelmann,~C.~B.; Courtois,~H.; Pekola,~J.~P. Thermal Conductance of a Single-Electron Transistor. \emph{Phys. Rev. Lett.} \textbf{2017}, \emph{119}, 077701\relax
\mciteBstWouldAddEndPuncttrue
\mciteSetBstMidEndSepPunct{\mcitedefaultmidpunct}
{\mcitedefaultendpunct}{\mcitedefaultseppunct}\relax
\EndOfBibitem
\bibitem[Scheibner \latin{et~al.}(2005)Scheibner, Buhmann, Reuter, Kiselev, and Molenkamp]{Scheibner05}
Scheibner,~R.; Buhmann,~H.; Reuter,~D.; Kiselev,~M.~N.; Molenkamp,~L.~W. Thermopower of a Kondo Spin-Correlated Quantum Dot. \emph{Phys. Rev. Lett.} \textbf{2005}, \emph{95}, 176602\relax
\mciteBstWouldAddEndPuncttrue
\mciteSetBstMidEndSepPunct{\mcitedefaultmidpunct}
{\mcitedefaultendpunct}{\mcitedefaultseppunct}\relax
\EndOfBibitem
\bibitem[Dutta \latin{et~al.}(2019)Dutta, Majidi, García~Corral, Erdman, Florens, Costi, Courtois, and Winkelmann]{Dutta2019}
Dutta,~B.; Majidi,~D.; García~Corral,~A.; Erdman,~P.~A.; Florens,~S.; Costi,~T.~A.; Courtois,~H.; Winkelmann,~C.~B. Direct Probe of the Seebeck Coefficient in a Kondo-Correlated Single-Quantum-Dot Transistor. \emph{Nano Letters} \textbf{2019}, \emph{19}, 506--511\relax
\mciteBstWouldAddEndPuncttrue
\mciteSetBstMidEndSepPunct{\mcitedefaultmidpunct}
{\mcitedefaultendpunct}{\mcitedefaultseppunct}\relax
\EndOfBibitem
\bibitem[Costi and Zlati\ifmmode~\acute{c}\else \'{c}\fi{}(2010)Costi, and Zlati\ifmmode~\acute{c}\else \'{c}\fi{}]{CostiZlati}
Costi,~T.~A.; Zlati\ifmmode~\acute{c}\else \'{c}\fi{},~V. Thermoelectric transport through strongly correlated quantum dots. \emph{Phys. Rev. B} \textbf{2010}, \emph{81}, 235127\relax
\mciteBstWouldAddEndPuncttrue
\mciteSetBstMidEndSepPunct{\mcitedefaultmidpunct}
{\mcitedefaultendpunct}{\mcitedefaultseppunct}\relax
\EndOfBibitem
\bibitem[Dong and Lei(2002)Dong, and Lei]{DongLei}
Dong,~B.; Lei,~X.~L. Effect of the Kondo correlation on the thermopower in a quantum dot. \emph{Journal of Physics: Condensed Matter} \textbf{2002}, \emph{14}, 11747\relax
\mciteBstWouldAddEndPuncttrue
\mciteSetBstMidEndSepPunct{\mcitedefaultmidpunct}
{\mcitedefaultendpunct}{\mcitedefaultseppunct}\relax
\EndOfBibitem
\bibitem[Gehring \latin{et~al.}(2019)Gehring, van~der Star, Evangeli, Le~Roy, Bogani, Kolosov, and van~der Zant]{Gehring_effheating}
Gehring,~P.; van~der Star,~M.; Evangeli,~C.; Le~Roy,~J.~J.; Bogani,~L.; Kolosov,~O.~V.; van~der Zant,~H. S.~J. {Efficient heating of single-molecule junctions for thermoelectric studies at cryogenic temperatures}. \emph{Applied Physics Letters} \textbf{2019}, \emph{115}, 073103\relax
\mciteBstWouldAddEndPuncttrue
\mciteSetBstMidEndSepPunct{\mcitedefaultmidpunct}
{\mcitedefaultendpunct}{\mcitedefaultseppunct}\relax
\EndOfBibitem
\bibitem[Frisenda \latin{et~al.}(2015)Frisenda, Tarku{\c c}, Gal{\'a}n, Perrin, Eelkema, Grozema, and van~der Zant]{Frisenda_different_anchors}
Frisenda,~R.; Tarku{\c c},~S.; Gal{\'a}n,~E.; Perrin,~M.~L.; Eelkema,~R.; Grozema,~F.~C.; van~der Zant,~H. S.~J. Electrical properties and mechanical stability of anchoring groups for single-molecule electronics. \emph{Beilstein J Nanotechnol} \textbf{2015}, \emph{6}, 1558--1567\relax
\mciteBstWouldAddEndPuncttrue
\mciteSetBstMidEndSepPunct{\mcitedefaultmidpunct}
{\mcitedefaultendpunct}{\mcitedefaultseppunct}\relax
\EndOfBibitem
\bibitem[Baum(2024)]{Thomas_Baum_thesis}
Baum,~T. Electronic transport signatures of two-electron interactions in all-organic single-molecule junctions. PhD. thesis, TU Delft, Delft, The Netherlands, 2024\relax
\mciteBstWouldAddEndPuncttrue
\mciteSetBstMidEndSepPunct{\mcitedefaultmidpunct}
{\mcitedefaultendpunct}{\mcitedefaultseppunct}\relax
\EndOfBibitem
\bibitem[Meir \latin{et~al.}(1993)Meir, Wingreen, and Lee]{Lee_oldAndersonmodel}
Meir,~Y.; Wingreen,~N.~S.; Lee,~P.~A. Low-temperature transport through a quantum dot: The Anderson model out of equilibrium. \emph{Phys. Rev. Lett.} \textbf{1993}, \emph{70}, 2601--2604\relax
\mciteBstWouldAddEndPuncttrue
\mciteSetBstMidEndSepPunct{\mcitedefaultmidpunct}
{\mcitedefaultendpunct}{\mcitedefaultseppunct}\relax
\EndOfBibitem
\bibitem[Nagaoka \latin{et~al.}(2002)Nagaoka, Jamneala, Grobis, and Crommie]{Nagaoka_Kondotemp}
Nagaoka,~K.; Jamneala,~T.; Grobis,~M.; Crommie,~M.~F. Temperature Dependence of a Single Kondo Impurity. \emph{Phys. Rev. Lett.} \textbf{2002}, \emph{88}, 077205\relax
\mciteBstWouldAddEndPuncttrue
\mciteSetBstMidEndSepPunct{\mcitedefaultmidpunct}
{\mcitedefaultendpunct}{\mcitedefaultseppunct}\relax
\EndOfBibitem
\end{mcitethebibliography}

\end{document}